\documentclass[journal,twoside,web]{ieeecolor}
\usepackage{lcsys}
\usepackage{cite}
\usepackage{amsmath,amssymb,amsfonts}
\usepackage{algorithmic}
\usepackage{graphicx}
\usepackage{textcomp}
\usepackage{amssymb}
\usepackage{amsmath}
\usepackage{algorithm}
\def\overbigdot#1{\overset{\hbox{\tiny$\bullet$}}{#1}}
 \newcommand{\tabitem}{~~\llap{\textbullet}~~}  
\def\BibTeX{{\rm B\kern-.05em{\sc i\kern-.025em b}\kern-.08em
    T\kern-.1667em\lower.7ex\hbox{E}\kern-.125emX}}
\begin{document}
\title{A Consensus-Based Generalized Multi-Population Aggregative Game with Application to Charging Coordination of Electric Vehicles}
\author{Mahsa Ghavami, Babak Ghaffarzadeh Bakhshayesh, Mohammad Haeri, \IEEEmembership{Senior Member, IEEE}, Giacomo Como, \IEEEmembership{Member, IEEE}, and Hamed Kebriaei, \IEEEmembership{Senior Member, IEEE}
\thanks{Manuscript received July 13, 2023; revised September 14, 2023; accepted October 7, 2023. Accepted by Editor in Cheif M.E. Valcher.The research is financially supported by the Iran National Science Foundation (INSF) under Grant 4021352}
\thanks{M. Ghavami and M. Haeri are with the Advanced Control Systems Lab, Department of Electrical Engineering, Sharif University of Technology, Tehran, Iran (e-mails: mahsa.ghavami@ee.sharif.edu and haeri@sharif.ir).
}
\thanks{B. Ghaffarzadeh Bakhshayesh and H. Kebriaei are with School of Electrical and Computer Engineering, College of Engineering, University of Tehran, Tehran, Iran (e-mails: b.ghaffarzadeh@ut.ac.ir and kebriaei@ut.ac.ir).}
\thanks{G. Como is with the Department of Mathematical Sciences, Politecnico di Torino, Torino, Italy (e-mail: giacomo.como@polito.it).}}

\maketitle
\thispagestyle{empty} % Removes the page number in the first page

\begin{abstract}
This paper introduces a consensus-based generalized multi-population aggregative game coordination approach with application to electric vehicles charging under transmission line constraints. The algorithm enables agents to seek an equilibrium solution while considering the limited infrastructure capacities that impose coupling constraints among the users. The Nash-seeking algorithm consists of two interrelated iterations. In the upper layer, population coordinators collaborate for a distributed estimation of the coupling aggregate term in the agents' cost function and the associated Lagrange multiplier of the coupling constraint, transmitting the latest updated values to their population's agents. In the lower layer, each agent updates its best response based on the most recent information received and communicates it back to its population coordinator. For the case when the agents' best response mappings are non-expansive,  we prove the algorithm's convergence to the generalized Nash equilibrium point of the game. Simulation results demonstrate the algorithm's effectiveness in achieving equilibrium in the presence of a coupling constraint.
\end{abstract}

\vspace{-.5 cm}
\section{Introduction}
\IEEEPARstart{I}{n} today's digital era, data availability is crucial for efficient decision-making and resource allocation in societal-scale systems, such as transportation electrification \cite{alleyne2023control}. However, integrating various data sources and coordinating their use presents challenges regarding accuracy and reliability. For instance, multiple online platforms like ChargeHub collect and distribute data on users' Electric Vehicle (EV) charging. The involvement of multiple platforms, each collecting data from a subset of users, can lead to suboptimal decisions and inefficient resource allocation in systems with limited capacity \cite{broo2020towards}.

To address these challenges, the study of multi-agent systems and game theory has gained significant interest, particularly in large-scale systems \cite{marden2018game}. Aggregative games with coupling constraints have emerged as a notable research area within this discipline. These games model competitive situations where actors experience aggregate effects from the entire population rather than direct interactions with individual agents while satisfying critical infrastructure constraints \cite{paccagnan2018nash}. Applications of generalized aggregative games include peer-to-peer energy markets \cite{belgioioso2022operationally} and coordination of EVs \cite{bakhshayesh2021decentralized}, \cite{10024972}.

Interestingly, the aggregative structure has been employed to address computational complexity in scenarios involving large populations. Proposed solution algorithms primarily operate in a non-centralized manner, utilizing decentralized and distributed algorithms \cite{paccagnan2018nash} and \cite{grammatico2017dynamic, belgioioso2017semi, de2018continuous}. In decentralized algorithms, agents do not communicate directly with each other but rely on a central coordinator that aggregates local decisions and broadcasts signals, such as dual variables, to all agents. Distributed algorithms, on the other hand, involve agents obtaining information through communication with their neighbors via a communication graph \cite{koshal2016distributed, belgioioso2020distributed, gadjov2020single}. These approaches show promise in addressing the challenges of large-scale multi-agent systems and coupling constraints.

In decentralized generalized Nash equilibrium-seeking algorithms, the authors in \cite{grammatico2017dynamic}, \cite{belgioioso2017semi} propose a forward-backward algorithm designed to identify a generalized Nash equilibrium within a strongly monotone game setting where the coordinator sends dual variables to all participants. Alternatively, in \cite{paccagnan2018nash}, a gradient-type algorithm is used to exhibit global convergence with fixed step sizes, assuming strong monotonicity in the pseudo-gradient of the game. The coordinator sends both the aggregate and updated dual variables in this scenario. Similarly, \cite{de2018continuous} presents a continuous-time variant of the algorithm, contingent upon strict monotonicity in the pseudo-gradient.

In the domain of distributed computation for the Nash equilibrium in aggregative games, the authors in \cite{koshal2016distributed} have proposed a dynamic average tracking scheme to estimate the unknown average aggregate in a distributed manner. Another study tackled coupling constraints using a monotone operator splitting method and the Krasnosel’skii-Mann fixed-point iterations \cite{belgioioso2020distributed}. Furthermore, aggregative games with coupling constraints have been explored using the forward-backward operator splitting technique \cite{gadjov2020single}. The survey \cite{ye2023distributed} examines recent equilibrium-seeking algorithms and their characteristics.

Existing research \cite{paccagnan2018nash} and \cite{grammatico2017dynamic, belgioioso2017semi, de2018continuous, koshal2016distributed, belgioioso2020distributed, gadjov2020single, ye2023distributed, benenati2023semi} has primarily focused on decentralized and distributed algorithms for large-scale multi-agent systems with a single population of users. However, these studies have often assumed that users receive information through a single coordinator or exchange it with neighboring agents. In today's real-world applications, we frequently encounter multiple coordinators, each responsible for a subset of users. While systems with a single coordinator allow for centralized control and straightforward data aggregation, multiple coordinators introduce challenges such as potential conflicting directives and the need for synchronization. This paper addresses the decentralized multi-population scenario where coordinators have only local data. Specifically, we develop a novel algorithm based on Krasnosel’skii-Mann iteration and consensus protocols. Rigorous analysis shows the equilibrium point of each local population mapping converges to the centralized generalized Nash equilibrium despite limited information. This significantly advances decentralized equilibrium computation for interconnected multi-agent systems.

The authors of \cite{kebriaei2021multipopulation} have investigated equilibrium seeking in multi-population aggregative games with decoupled strategy sets, where each agent's feasible strategy set is not impacted by other agents. Nevertheless, motivated by the importance of shared constraints in energy management applications, the current paper takes a step further by examining coupling constraints among users in a multi-population game. 

This paper aims to develop a mechanism that coordinates information from multiple platforms, considering the coupling constraints imposed by user data, to improve decision-making and resource allocation in infrastructure systems with limited capacities. Specifically, we focus on decentralized equilibrium-seeking methods in multi-population aggregative games with coupling constraints. By tackling this research gap, our consensus-based algorithm enables Population Coordinators (PCs) to estimate global aggregates through local estimates, thereby extending the existing literature and offering new insights into this domain. The algorithm allows players to optimize their cost functions based on local estimates and operates through a two-level framework. At the top level, PCs exchange information with neighboring coordinators, while at the bottom level, individual players make decisions independently. The algorithm aims to compute the Nash equilibrium, considering the interdependencies among players and the shared constraints. The contributions of this paper can be summarized as:
\begin{itemize}
\item Extension of the results on multi-population aggregative game \cite{kebriaei2021multipopulation} to incorporate coupling constraints arising from limited energy or infrastructure resources.
\item Development of a semi-decentralized equilibrium-seeking algorithm in which the coupling aggregate term of the cost functions and Lagrange multiplier of the coupling constraint are distributively estimated by local coordinators of each population.
\item Providing the proof of convergence of the proposed Nash-seeking algorithm that consists of two interrelated iterations of PCs and agents.
\end{itemize}

\smallskip{
\vspace{-.2 cm}
\subsubsection*{Notation}
$\mathbb{N}, \mathbb{R}, \mathbb{R}_{>0}$, and $\mathbb{R}_{\geq 0}$ are sets of natural, real, positive, and non-negative numbers. $\mathbb{S}^n$ denotes the set of symmetric $n\times n$ matrices. $x^{T}, \|x\|$, and $\|x\|_{P}:=\sqrt{x^TPx}$ denote transpose, infinity norm, and induced norm of vector $x$. $\|B\|$ is the induced matrix infinity norm, simply the maximum absolute column sum of matrix $B$. Given vectors $x_1,...,x_L$, the column augmented vector $x_l$ is defined as $\mathrm{col}(x_1,...,x_L)=[x_1^T,...,x_L^T]^T$.  $I_n$ denotes the identity matrix and $\textbf{1}_n$ is a vector with all elements one. The notation $K\succ0$ ($K\succeq0$) denotes that $K$ is symmetric and has positive (non-negative) eigenvalues.}
\vspace{-.3 cm}
\section{Multi-Population Aggregative Game with Coupling Constraints}
\vspace{-.3 cm}
\subsection{Game Setup}
%\subsection{Selecting a Template (Heading 2)}
We consider a set of $L$ populations of agents, each of which has a PC
%population coordinator has been abbreviated to PC
$l \in\mathcal{L}=\{1,...,L\}$ and $N_{l}$ agents whose set is denoted by $\mathcal{N}_{l}$. Thus, the total number of agents is $N=\sum_{l\in\mathcal{L}}N_{l}$. We assume that PCs can exchange information through a time-varying directed graph $G(\mathcal{L},\mathcal{E}^k)$, where $\mathcal{E}^k$ is the set of directed edges at time $k$. Here, $(l,l')\in\mathcal{E}^k$ means that PC $l$ can receive information from PC $l'$ and
$W^{k}\in\mathbb{R}^{L\times{L}}$ represents a weight matrix of the communication graph $G$ at time $k$ so that $w_{l,l'}^{k}$ is a weight that PC $l$ assigns to the information coming from PC $l'$ at time $k$. If $(l,l')\in\mathcal{E}^{k}$ , $w_{l,l'}^{k}>0$, otherwise $w_{l,l'}^{k}=0$.
The set of neighbors of PC $l$ at time $k$ consists  of all PCs from which PC $l$ receives information and is denoted by $\mathcal{L}_{l}^{k}=\{l^{'}\in\mathcal{L}|(l,l^{'})\in\mathcal{E}^k\}$.
Each agent $i$ in population $l$ chooses strategy $x_{l,i}$ according to its individual constraint set $\mathcal{X}_{l,i} \subset\mathbb{R}^{n}$ and a linear coupling constraint as 
 \begin{equation}\label{aggregative_term}
 \sigma\in\mathcal{C},
\end{equation}
where $\sigma:=\frac{1}{L}\sum_{l \in \mathcal{L}}\sum_{i\in\mathcal{N}_{l}}\delta_{l,i}x_{l,i}$ is an aggregate term. $\delta_{l,i}$ are non-negative coefficients such that $\sum_{i\in\mathcal{N}_l}\delta_{l,i}=1$ and are determined by PC $l$. We further assume $\mathcal{C}\subseteq\frac{1}{L}\sum_{l\in\mathcal{L}}\sum_{i\in\mathcal{N}_{l}}\delta_{l,i}\mathcal{X}_{l,i}\subset\mathbb{R}^n$
and $\mathcal{X}_{l,i}$ and $\mathcal{C}$ are compact and convex subsets of $\mathbb{R}^n$ and satisfy Slater's constraint qualification.
For convenience, We define $x_{l}:=$col$(x_{l,1},...,x_{l,N_{l}})$ and $x:=$col$(x_{1},...,x_{L})$.

The cost function of each agent $i$ is defined as
\begin{equation}\label{cost_function}
J_{l,i}(x_{l,i},\sigma,\lambda):=f_{l,i}(x_{l,i})+(C\sigma+K\lambda)^{T}x_{l,i},
\end{equation}
where $\lambda\in\mathbb{R}^{n}$ is a control vector to penalize the violation of the coupling constraint. $f_{l,i}:\mathbb{R}^{n}\to \mathbb{R}$ is a continuous strongly convex function and $C\in \mathbb{S}^{n}$ is a given weight matrix 
%{that illustrates the effects of the average among strategies of all agents} 
while the invertible matrix $K\in\mathbb{S}^{n}$ is a design choice that could be used to guarantee the convergence of the decentralized learning algorithm. Let $\sigma_l$ and $\lambda_{l}$ be the local estimates of $\sigma$ and $\lambda$ in \eqref{cost_function}, which are determined by PC $l$ \eqref{cost_function} depending on his own strategy $x_{l,i}\in\mathcal{X}_{l,i}$ and on the strategy of other agents through the aggregate term $\sigma$, thus we have an aggregative game. 
The cost function of each agent is also influenced by the unique control vector $\lambda$. %Therefore, we have a competitive aggregative game. 
\newtheorem{theorem1}{Remark} 
\begin{theorem1}
The proposed aggregative game with coupling constraint \eqref{aggregative_term} is applicable to the management of self-interested agents in energy systems that are coupled in the aggregated form like %Applications include 
the coordination of EV charging with transmission line constraints \cite{Ghavami2013Price}, traffic and charging stations with infrastructure limitations such as road capacity and electric power resources \cite{bakhshayesh2021decentralized}, and peak shaving for residential energy storage systems \cite{Joshi2021Decentralized}. %The application of EV charging coordination with transmission line constraints is studied in Section V.
\end{theorem1} 

Generalized multi-population Nash equilibrium is a set of strategies in which no agent could benefit from unilaterally deviating from its own strategy and the coupling constraints are satisfied. According to [\citenum{grammatico2017dynamic}, Remark 2], the $\epsilon$-Nash equilibrium point of the proposed generalized multi-population aggregative game is defined as follows.

\newtheorem{theorem}{Definition}  
\begin{theorem}
A pair $(\hat{x},\hat{\lambda})$ is a generalized multi-population $\epsilon$-Nash equilibrium with $\epsilon>0$ for agents with cost \eqref{cost_function} and coupling constraint \eqref{aggregative_term} if $\hat{\sigma}:=\frac{1}{L}\sum_{l\in\mathcal{L}}\sum_{i\in\mathcal{N}_l}\delta_{l,i}\hat{x}_{l,i}\in\mathcal{C}$ and for all $l\in\mathcal{L}$, $i\in\mathcal{N}_l$, and $r\in \mathcal{X}_{l,i}$, we have 
\begin{equation}\label{def}
  \begin{split}
&J_{l,i}(\hat{x}_{l,i},\hat{\sigma},\hat{\lambda})\leq \epsilon+\min_{r\in\mathcal{X}_{l,i}}J_{l,i}\bigg(r,\frac{1}{L}\delta_{l,i}r+\\&\frac{1}{L}\sum\limits_{i^{'}\in \mathcal{N}_{l}\backslash i}\delta_{l,i'}\hat{x}_{l,i^{'}}+\frac{1}{L}\sum\limits_{l^{'}\in\mathcal{L}\backslash l}\sum\limits_{i^{'}\in\mathcal{N}_{l^{'}}}\delta_{l',i'}\hat{x}_{l^{'},i^{'}},\hat{\lambda}\bigg). 
\end{split} %\tag{3} %\setcounter{equation}{3}
\end{equation}
\end{theorem} 
$(\hat{x},\hat{\lambda})$ is a generalized multi-population Nash equilibrium if \eqref{def} holds with $\epsilon=0$. At the end of this paper, we have shown that our proposed algorithm converges to the multi-population $\epsilon$-Nash equilibrium where $\epsilon$ converges to zero in the limit of infinite total population size.

We define the agent's best response mapping $x_{l,i}^*:\mathbb{R}^{n}\to\mathcal{X}_{l,i}$ to the incentive signal $v=C\sigma+K\lambda$ as 
\begin{equation}\label{best_response}
    x_{l,i}^*(v):=\arg\min_{u\in\mathcal{X}_{l,i}}f_{l,i}(u)+v^Tu.
\end{equation}

By defining $y:=\mathrm{col}(\sigma,\lambda)$, we group together the optimal response mappings into the local and global aggregation mapping $\mathcal{A}_l$ and $\mathcal{A}$ as 
\begin{equation}\label{A_l}
    \mathcal{A}_l(y):=\sum_{i\in\mathcal{N}_l}\delta_{l,i}x_{l,i}^*(v),\quad \mathcal{A}(y):=\frac{1}{L}\sum_{l\in\mathcal{L}}\mathcal{A}_l(y).
\end{equation}

Now, results in \cite{grammatico2017dynamic} can be used to establish the equivalence between generalized Nash equilibrium and a fixed point of a certain mapping. Specifically, for a symmetric matrix $K$, define
\begin{equation}\label{x0*}
    x^{**}(y):=\arg\min_{z\in\mathcal{C}}\frac{1}{2}z^Tz+(K(\sigma-\lambda))^Tz,
\end{equation}
and let $\mathcal{T}:\mathbb{R}^{2n}\to\mathbb{R}^{2n}$
\begin{equation}\label{T}
\mathcal{T}(y):=B(y-\eta\Gamma(y)), \Gamma(y):=-\begin{bmatrix}
\mathcal{A}(y)\\
2\mathcal{A}(y)-x^{**}(y)
\end{bmatrix},
\end{equation}
where $B=(I+\eta M)^{-1}, M=\begin{bmatrix}I_n & 0\\I_n & 0\end{bmatrix}$. Then, we have the following result.

\newtheorem{theorem33}{Lemma}
\begin{theorem33}
If $K\succ0$ and $C+K\succ0$, then for sufficiently small constant $\eta>0$, the mapping $\mathcal{T}(.)$ in \eqref{T} is non-expansive, it admits a unique fixed point $\hat{y}=\mathcal{T}(\hat{y})$ and it is the generalized Nash equilibrium of the game for large number of agents.
\end{theorem33}
Lemma 1 follows from [\citenum{grammatico2017dynamic}, Proof of Theorem 1 and 4] as explained in detail in APPENDIX I. 
Although the fixed point of mapping $\mathcal{T}$ in \eqref{T} is also equivalent to the generalized Nash equilibrium for our multi-population case, the equilibrium-seeking algorithm proposed in \cite{grammatico2017dynamic} can no longer be used since the aggregate term $\sigma$ and $\lambda$ is controlled by multi-coordinators connected through $G(\mathcal{L},\mathcal{E}^k)$.
To determine the fixed point of mapping $\mathcal{T}$ in our multi-population case, we define mapping $\mathcal{T}_l:\mathbb{R}^{2n}\to\mathbb{R}^{2n}$ and $\Gamma_l:\mathbb{R}^{2n}\to\mathbb{R}^{2n}$ as follows.
\begin{equation}\label{Tl}
 \mathcal{T}_l(y):=B(y-\eta\Gamma_l(y)), \Gamma_l(y):=-\begin{bmatrix}
\mathcal{A}_l(y)\\
2\mathcal{A}_l(y)-x^{**}(y)
\end{bmatrix}.
\end{equation}

\begin{figure}[thpb]  
      \centering
          \includegraphics[scale=0.46]{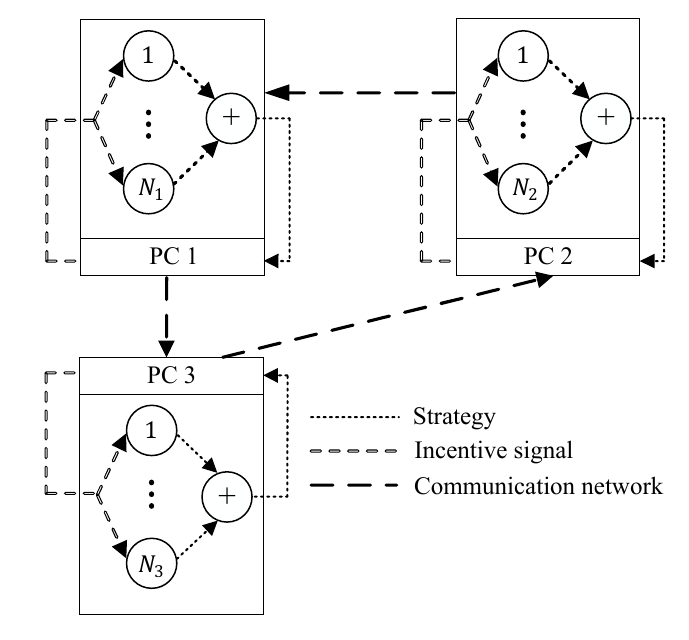} 
          \vspace{-.3 cm}
      \caption{Schematic of information flow of a three-population network. }
     
      \label{figurelabel1}
\end{figure}

According to (5), since $\mathcal{A}(y)=\frac{1}{L}\sum_{l\in\mathcal{L}}\mathcal{A}_l(y)$, it is easy to show that $\Gamma(y)=\frac{1}{L}\sum_{l\in\mathcal{L}}\Gamma_{l}(y)$. Then, $\mathcal{T}(y)=\frac{1}{L}\sum_{l\in\mathcal{L}}\mathcal{T}_l(y)$.
  We aim to design a consensus-based decentralized algorithm such that the local signals $y_l:=\mathrm{col}(\sigma_l,\lambda_l)$ converge to the fixed point of mapping $\mathcal{T}$ in \eqref{T} as the generalized $\epsilon$-Nash equilibrium of the game. The schematic of the information flow is illustrated in Fig. \ref{figurelabel1}. Each PC broadcasts an incentive signal to its population. Then, each agent updates its strategy and sends it back to its PC. Based on local agents' strategies and communication with neighbors, each PC updates the incentive signal to be broadcast at the next iteration. 
  \vspace{-.3 cm}
\vspace{-.9 cm}
\subsection{Consensus-Based Decentralized Equilibrium Seeking}
To develop an equilibrium-seeking algorithm with lightweight information exchange, we assume that the agents have no information about the parameters of their cost functions, including $C$ and $K$ in \eqref{cost_function}, the structure of aggregate term $\sigma$ and parameters $\delta_{l,i}$, control vector $\lambda$, and strategies of other agents. Instead, we assume that each agent responds optimally to the local incentive signals $v_l:=C\sigma_l+K\lambda_l$, which are broadcasted by PC $l$. 
The proposed algorithm is summarized in Algorithm 1: At iteration $k$, each PC sends $v_{l}^k$, which is a function of the local estimation $y_l^k$, to its population. Afterwards, each agent reacts optimally to $v_l^k$ by $x_{l,i}^*(v_l^k)$ in \eqref{best_response}. Then, having the local aggregate strategies  and $x^{**}(y_l^k)$, PC $l$ updates the output of $\mathcal{T}_l$ as $\mathcal{T}_l^k:=\mathcal{T}_l(y_l^k)$.
Additionally, PC $l$ receives the local incentive signal of its neighbors through the communication network and updates $y_l^k$ based on a combination of Krasnoselskij-Mann (KM) iteration \cite{Borwein1992Krasnoselski} and consensus update as follows
\begin{equation}\label{y_lk+1}
    y_l^{k+1}=(1-\alpha_k)(\sum_{l'\in\mathcal{L}}
w_{l,l'}^ky_{l'}^k)+\alpha_k\mathcal{T}_l^k,
\end{equation} 
where $\alpha_k\in(0,1)$, $k\in\mathbb{N}$ is an step size.
The convergence analysis of Algorithm 1 is challenging since the consensus update and aggregate term update are performed consecutively by each coordinator, which introduces a coupling between these two iterations.
\vspace{-.2 cm}
\begin{table}[h]
%\caption{An Example of a Table}
\label{table_example}
\begin{center}
\begin{tabular}{ l }
\hline
\textbf{Algorithm 1} Consensus-based decentralized equilibrium seeking\\
\hline
\textbf{Initialization:} $k\leftarrow 1$, $y_l^1\leftarrow y_l^0, \forall l\in \mathcal{L}$\\ $(y_l^0=\mathrm{col}(\sigma_l^0,\lambda_l^0)\in\mathcal{C}\times\mathbb{R}^n)$\\
\textbf{Iteration k:} \\
\begin{tabular}{|l}
\tabitem Each PC $l\in\mathcal{L}$ broadcasts
 $v_l^k=C\sigma_l^k+K\lambda_l^k$
to\\ its population and computes $x^{**}(y_l^k)$ from \eqref{x0*}.\\
\tabitem For each $l\in\mathcal{L}$ and $i\in\mathcal{N}_l$, the agents compute in parallel \\$x_{l,i}^*(v_l)$ as \eqref{best_response}.\\
\tabitem Each PC $l\in\mathcal{L}$ computes $\mathcal{T}_l$ from \eqref{Tl} and updates $y_l^k$\\ by using $\mathcal{T}_l$ and communicating with its neighbors:\\
$y_l^{k+1}=(1-\alpha_k)(\sum_{l'\in\mathcal{L}}w_{l,l'}^ky_l^k)+\alpha_k\mathcal{T}_l(y_l^k)$
\end{tabular}
\\
$k\leftarrow k+1$\\
\hline
\vspace{-.68 cm}
\end{tabular}
\end{center}
\end{table}
\vspace{-.5 cm}
\section{Convergence Analysis}
\vspace{-.1 cm}
In this section, we investigate the convergence of Algorithm 1.  First, we prove the convergence of the local incentive signals, denoted as $y_l^k$ to the average signal $\overline{y}^k$, defined as the mean of all $y_l^k$ values

\begin{equation}\label{yb}
    \overline{y}^k:=\frac{1}{L}\sum_{l\in\mathcal{L}}y_l^k.
\end{equation}
Next, we establish the convergence of $\overline{y}^k$ to the fixed point of the mapping $\mathcal{T}$ defined in \eqref{T}, which represents the generalized Nash equilibrium of problem \eqref{cost_function}.
% To begin, we establish the convergence of each $y_l^k$ to $\overline{y}^k$ as $k\to \infty$. %This convergence result is based on the following two assumptions:

\newtheorem{theorem2}{Assumption}
\begin{theorem2}
The sequence of $\alpha_k>0$, $k\in\mathbb{N}$ is non-increasing, non-summable,
%$\sum_k\alpha_k=\infty$,
and square-summable.% $\sum_k(\alpha_k)^2<\infty$.
\end{theorem2}

\begin{theorem2} $\forall l,l'\in\mathcal{L}$ and $k\in\mathbb{N}$, (1): There exists a scalar $0<\mu<1$ such that, $w_{l,l}^{k}>\mu$ %for all $l\in\mathcal{L}$ and $k\in\mathbb{N}$
and $w_{l,l'}^{k}>\mu$ if $w_{l,l'}^{k}>0$. (2): $W^{k}$ is doubly stochastic which means, $\sum_{l'=1}^{L}w_{l,l'}^{k}=1$ %for all $l\in\mathcal{L}$ and 
%$k\in\mathbb{N}$ and 
and $\sum_{l=1}^{L}w_{l,l'}^{k}=1$. %for all $l'\in\mathcal{L}$ and $k\in\mathbb{N}$
(3): There is an integer $\overline{k}\geq1$ such that  $G(\mathcal{L},\cup_{k^{'}=1,...,\overline{k}}\mathcal{E}^{k+k^{'}})$ is strongly connected.
\end{theorem2}
 By averaging $y_l^k$ in \eqref{y_lk+1} over the entire population, utilizing \eqref{yb}, and considering Assumption 2, we obtain the following update equation for $\overline{y}^{k+1}$
\begin{equation}\label{ybk}
    \overline{y}^{k+1}=(1-\alpha_k)\overline{y}^{k}+\frac{1}{L}\alpha_k\sum_{l\in\mathcal{L}}\mathcal{T}_l^k.
\end{equation}
Our objective is to prove the consensus of the local variables $y_l^k$ towards $\overline{y}^k$.

\newtheorem{theorem3}{Theorem} 
\begin{theorem3}
Let Assumptions 1 and 2 hold, then $\lim_{k\to\infty}\max_{l\in\mathcal{L}}\|y_l^k-\overline{y}^k\|=0.$
\end{theorem3}
%%%%%%%%%%%%%you have discuss design choice P
\begin{proof}
We define the transition matrix $\phi^{k-1,0}:=W^{k-1}W^{k-2}...W^{0}$ and $\phi^{k,k}:=W^k, \forall k\in\mathbb{N}$, where $\big[\phi^{k_2,k_1}\big]_{l,l'}$ denotes the $(l,l')$ element of the matrix $\phi^{k_2,k_1}$. By relating the local estimations in \eqref{y_lk+1} from $0$ to $k$, $y_l^{k}$ is derived as
\begin{equation}\label{ylk}
  \begin{split}
y_l^k=\alpha_{k-1}\mathcal{T}_l^{k-1}&+\sum_{k_1=0}^{k-2}\sum_{l'\in\mathcal{L}}\big[\hat{\phi}^{k-1,k_1+1}\big]_{l,l'}\alpha_{k_1}\mathcal{T}_{l'}^{k_1}\\&+\sum_{l'\in\mathcal{L}}\big[\hat{\phi}^{k-1,0}\big]_{l,l'}y_{l'}^0,
\end{split}
\end{equation}
where $\hat{\phi}^{k_2,k_1}=\phi^{k_2,k_1}F^{k_2,k_1}$ and $F^{k_2,k_1}=(1-\alpha_{k_2})...(1-\alpha_{k_1})$. Similarly, by relating the average signal in \eqref{ybk} from 0 to $k$, $\overline{y}^{k}$ can be computed as
\begin{align}\label{ybk14}
&\overline{y}^k=\frac{1}{L}\alpha_{k-1}\sum_{l\in\mathcal{L}}\mathcal{T}_l^{k-1}+\frac{1}{L}\sum_{k_1=0}^{k-2}\sum_{l\in\mathcal{L}}F^{k-1,k_1+1}\alpha_{k_1}\mathcal{T}_{l}^{k_1}\nonumber\\
&+F^{k-1,0}\overline{y}^0.
\end{align}

Based on \eqref{ylk} - \eqref{ybk14}, the distance between $y_l^k$ to $\overline{y}^k$ bounds as follows
\begin{equation}\label{norm15}
\begin{split}
\|y_l^k-\overline{y}^k\|\leq \alpha_{k-1}\|\mathcal{T}_l^{k-1}\|+\frac{1}{L}\alpha_{k-1}\sum_{l\in\mathcal{L}}\|\mathcal{T}_l^{k-1}\|+\\\sum_{k_1=0}^{k-2}\sum_{l'\in\mathcal{L}}\bigg|\big[\phi^{k-1,k_1+1}\big]_{l,l'}-\frac{1}{L}\bigg|F^{k-1,k_1+1}\alpha_{k_1}\|\mathcal{T}_{l'}^{k_1}\|+\\\sum_{l'\in\mathcal{L}}\bigg|\big[\phi^{k-1,0}\big]_{l,l'}-\frac{1}{L}\bigg|F^{k-1,0}\|y_{l'}^0\|.
\end{split}
\end{equation}

Thanks to Assumption 2 and [\citenum{Nedic2009Distributed}, Prop. 1, b], we have
$$\bigg|[\phi^{k_2,k_1}]_{l,l'}-\frac{1}{L}\bigg|\leq c\gamma^{k_2-k_1}, \gamma\in(0,1), c>0.$$
Therefore, \eqref{norm15} can be written as
\begin{align}\label{norm16}
&\|y_l^k-\overline{y}^k\|\leq 2\alpha_{k-1}\max_{l\in\mathcal{L}}\|\mathcal{T}_l^{k-1}\|+Lc\gamma^{k-1}\max_{l\in\mathcal{L}}\|y_l^0\|+\nonumber\\
&Lc\sum_{k_1=0}^{k-2}\gamma^{k-k_1-2}\alpha_{k_1}\max_{l\in\mathcal{L}}\|\mathcal{T}_l^{k_1}\|.
\end{align}
We aim to determine the upper bound of $\|\mathcal{T}_l^k\|$ to obtain an upper bound for \eqref{norm16}. Since the feasible constraint set $\mathcal{X}_{l,i}$ and the coupling constraint set $\mathcal{C}$ is compact, $\Gamma_l(y_l^k)$ in \eqref{Tl} is bounded, i.e., there exists a constant $e\in\mathbb{R}_{>0}$ such that $\|\Gamma_l(y_l^k)\|\leq e$. Additionally, it is easy to show that $\|B\|=\frac{1}{1+\eta}$. Then, the upper bound for $\mathcal{T}_l$ in \eqref{Tl} is derived as follows. 
\begin{equation}\label{nTl}
    \|\mathcal{T}_l^k\|\leq\frac{1}{1+\eta}(\|y_l^k\|+\eta e)
\end{equation}

According to \eqref{nTl}, to prove boundedness of $\|\mathcal{T}_l^k\|$, it suffices to show that the sequence $\|y_l^k\|$ is bounded. Using proof by induction, we aim to prove that if there exists constant $\rho\in\mathbb{R}$ such that $\|y_l^k\|$ is bounded ($\|y_l^k\|\leq \rho$), then $\|y_l^{k+1}\|$ is bounded ($\|y_l^{k+1}\|\leq \rho$). 
Using \eqref{y_lk+1} and \eqref{nTl}, we  have
\begin{equation}\label{nylk}
        \|y_l^{k+1}\|\leq\rho
        +\frac{\alpha_k\eta}{1+\eta}(e-\rho). 
\end{equation}
If $\rho>e$, then $\|y_l^{k+1}\|\leq\rho$ and there exist a constant $\theta\in\mathbb{R}_{>0}$ such that $\|\mathcal{T}_l^k\|\leq\theta$. Therefore, \eqref{norm16} can be written as
\begin{equation}\label{eq19}
    \|y_l^k-\overline{y}^k\|\leq 2\alpha_{k-1}\theta+Lc\gamma^{k-1}\rho+Lc\theta\sum_{k_1=0}^{k-2}\gamma^{k-k_1-2}\alpha_{k_1}.
\end{equation}
Since $\alpha_k$ is square-summable, necessarily holds that $\lim_{k\to\infty}\alpha_k=0$ and the first term on the right side of \eqref{eq19} converge to zero. Also, owing to $\gamma\in(0,1)$, the second term converges to zero. It is easy to show that $\lim_{k\to\infty}\sum_{k_{1}=0}^{k}\gamma^{k-k_1}\alpha_{k_1}=0$, thus $\lim_{k\to\infty}\|y_l^k-\overline{y}^k\|=0$.
\end{proof}
Now, we aim to prove the local incentive signals converge to the fixed point of a mapping which is the generalized $\epsilon$-Nash equilibrium of \eqref{cost_function}.

\begin{theorem2}
The best response mapping $x_{l,i}^*(v_l), l\in\mathcal{L}, i\in\mathcal{N}_l$ in \eqref{best_response} is non-expansive. 
\end{theorem2}

\begin{theorem1}
Non-expansiveness of $x_{l,i}^*(v_l)$ depends on the choice of the function $f_{l,i}(.)$. For some classes of the functions, these sufficient conditions can be made explicitly, e.g. if $f_{l,i}(u)=u^{T}Qu+p^Tu$ with $Q\succ 0$ and $p\in\mathbb{R}^n$, then according to \cite{Grammatico2016Decentralized}, $x_{l,i}^*(v_l)$ is non-expansive whenever $Q-\frac{1}{4}Q^{-1}\succeq0$. In this case, $\mathcal{A}(.)$ in \eqref{A_l} is a convex combination of the best response mappings and thus is non-expansive.
\end{theorem1}

\begin{theorem3}
Under Assumptions 1-3, $\forall l\in\mathcal{L}$ and $k\in\mathbb{N}$, the sequence $y_l^k$ converges to the fixed point of mapping $\mathcal{T}$. 
\end{theorem3}
\begin{proof}  
Since the coupling constraint in \eqref{aggregative_term} is assumed linear and the objective function in \eqref{x0*} is quadratic, then 
   $\frac{1}{L}\sum_{l\in \mathcal{L}}x^{**}(y_l^k)=x^{**}(\overline{y}^k)$ (see \cite{Lecture}) and 
\begin{equation}\label{proof19}
    \frac{1}{L}\sum_{l\in\mathcal{L}}\mathcal{T}_l(y_l^k)=\mathcal{T}(\overline{y}^k)+E^k,
   E^k=B\eta\begin{bmatrix}
       -\mathcal{A}(\overline{y}^k)+\mathcal{A}(y_l^k)\\2(-\mathcal{A}(\overline{y}^k)+\mathcal{A}(y_l^k))
   \end{bmatrix}
\end{equation}
Therefore, $\overline{y}^{k+1}$ in \eqref{ybk} can be written as 
\begin{equation}\label{yb20}
    \overline{y}^{k+1}=(1-\alpha_k)\overline{y}^{k}+\alpha_k(\mathcal{T}(\overline{y}^k)+E^k).
\end{equation}

According to \cite{KIM2007Robust}, the sequence $\overline{y}^k, k\in\mathbb{N}$ converges to the fixed point of the mapping $\mathcal{T}$ if $\sum_{k=1}^{\infty}\alpha_k(1-\alpha_k)=\infty$ and $\sum_{k=1}^{\infty}\alpha_k\|E^k\|_P<\infty$. 

\begin{theorem1}
The non-expansiveness of the mapping $\mathcal{T}$ is a prerequisite to proving the robustness of $\overline{y}^{k}$ in \eqref{yb20}. The method to design $K$ and $x^{**}(.)$ to satisfy the non-expansiveness of mapping $\mathcal{T}$ in \eqref{T} is given in APPENDIX I. 
\end{theorem1}
Owing to Assumption 1, the first condition is satisfied. We only need to prove $\sum_{k=1}^{\infty}\alpha_k\|E^k\|_P<\infty$. 
We can write 
\begin{align}\label{proof20}
&\sum_{k=1}^{\infty}\alpha_k\|E^k\|_P\leq \hat{c}\sum_{k=1}^{\infty}\alpha_k\|\mathcal{A}(y_l^k)-\mathcal{A}(\overline{y}^k)\|\nonumber\\
&\leq \hat{c}\sum_{k=1}^{\infty}\alpha_k\|y_l^k-\overline{y}^k\|,
\end{align}
where $\hat{c}=\eta \sqrt{5\lambda_{\max}(B^TPB)}$. The second inequality in \eqref{proof20} has resulted from Assumption 3. Therefore, it suffices to prove that 
$\sum_{k=1}^{\infty}\alpha_k\|y_l^k-\overline{y}^k\|<\infty$.
According to \eqref{eq19},
\begin{equation}\label{proof21}
\begin{split}
&\sum_{k=1}^{\infty}\alpha_k\|y_l^k-\overline{y}^k\|\leq 2\theta\sum_{k=1}^{\infty}\alpha_k\alpha_{k-1}
+Lc\rho\sum_{k=1}^{\infty}\alpha_k\gamma^{k-1}+\\
&Lc\theta\sum_{k=1}^{\infty}\alpha_k\sum_{k_1=0}^{k-2}\gamma^{k-k_1-2}\alpha_{k_1}.
\vspace{-.3 cm}
\end{split}
\vspace{-.3 cm}
\end{equation}
Based on Assumption 1 and $\gamma\in(0,1)$, the first two terms on the right-hand side of \eqref{proof21} are bounded. As for the last term, since $\alpha_{k}$ is non-increasing and $\gamma\in(0,1)$
\begin{equation*}
  \begin{split}
\sum_{k=1}^{\infty}\alpha_k\sum_{k_1=0}^{k-2}\gamma^{k-k_1-2}\alpha_{k_1}\leq \sum_{k=1}^{\infty}\sum_{k_1=0}^{k-2}(\alpha_{k_1})^2\gamma^{k-k_1-2}
=\\\sum_{k_1=0}^{\infty}(\alpha_{k_1})^2\sum_{k=k_1+2}^{\infty}\gamma^{k-k_1-2}<\infty.
\end{split}
\end{equation*}
Therefore $\sum_{k=1}^{\infty}\alpha_k\|E^k\|<\infty$.
\end{proof}

\begin{theorem1}
Theorems 1 and 2 conclude that Algorithm 1 converges, meaning that the local incentive signals $y_l, l\in\mathcal{L}$ converge to a fixed point of mapping $\mathcal{T}$. While this analysis differs from single population case \cite{grammatico2017dynamic}, due to coupled consensus and aggregate terms update iterations, nevertheless, as for verifying the correspondence of the fixed point of mapping $\mathcal{T}$ and $\epsilon$-Nash equilibrium point, 
according to [\citenum{grammatico2017dynamic}, Theorem 1], it is straightforward to show that a set of strategies which are best responses to the fixed point of mapping $\mathcal{T}$ is $\epsilon$-Nash equilibrium. More details are given in APPENDIX II.
\end{theorem1}
\vspace{-.4 cm}
\section{Charging Control of EVs}
We consider the problem of the charging schedule of EVs when there are transmission line constraints. We assume that there are $L$ populations of EVs, each with a charging station coordinator. The charging station coordinators exchange information through a communication graph $G(\mathcal{L},\mathcal{E}^k)$ to estimate the aggregative charging demand of the whole population of EVs. Each EV $i\in\mathcal{N}_l$ aims to control its charging demand $x_{l,i}:= [x_{l,i}^t]_{t=1}^{n}$ over a charging interval $\mathcal{I}=\{1,...,n\}$ and to minimize its cost subject to the individual and coupling constraints which are described as follows.

The cost function of EV $i\in\mathcal{N}_l$ includes the battery degradation and charging costs and is given as 
\begin{equation}\label{ex24}
    J_{l,i}(x_{l,i},\sigma)=q_{l,i}x_{l,i}^Tx_{l,i}+p_{l,i}^{T}x_{l,i}+(a(\sigma+d)+b1_n)^Tx_{l,i},
\end{equation}
where $q_{l,i}>0$ and $p_{l,i}\in\mathbb{R}_{\geq 0}^n$ are  parameters of the battery degradation cost and $a>0$ and $b>0$ are parameters of the unit price function. Vector $d\in\mathbb{R}^n$ represents the normalized non-EV demand. The charging demand of EV $i\in\mathcal{N}_{l}$ at $t\in\mathcal{I}$ must satisfy
\begin{equation}\label{ex25}
    x_{l,i}\in\mathcal{X}_{l,i}=\{x_{l,i}^t|x_{l,i}^t\in[\underline{x}_{l,i},\overline{x}_{l,i}], 1^Tx_{l,i}=\beta_{l,i}\}.
\end{equation}
\vspace{-.2 cm}
Also, we consider transmission line constraints as
\begin{equation}\label{ex26}
    0\leq\frac{1}{L}\sum_{l\in\mathcal{L}}\sum_{i\in\mathcal{N}_l}\delta_{l,i}x_{l,i}^t\leq\overline{s}^t,\overline{s}^t\in\mathbb{R}_{\geq 0}.
\end{equation}

\begin{figure}[thpb]
      \centering
          \includegraphics[scale=0.42]{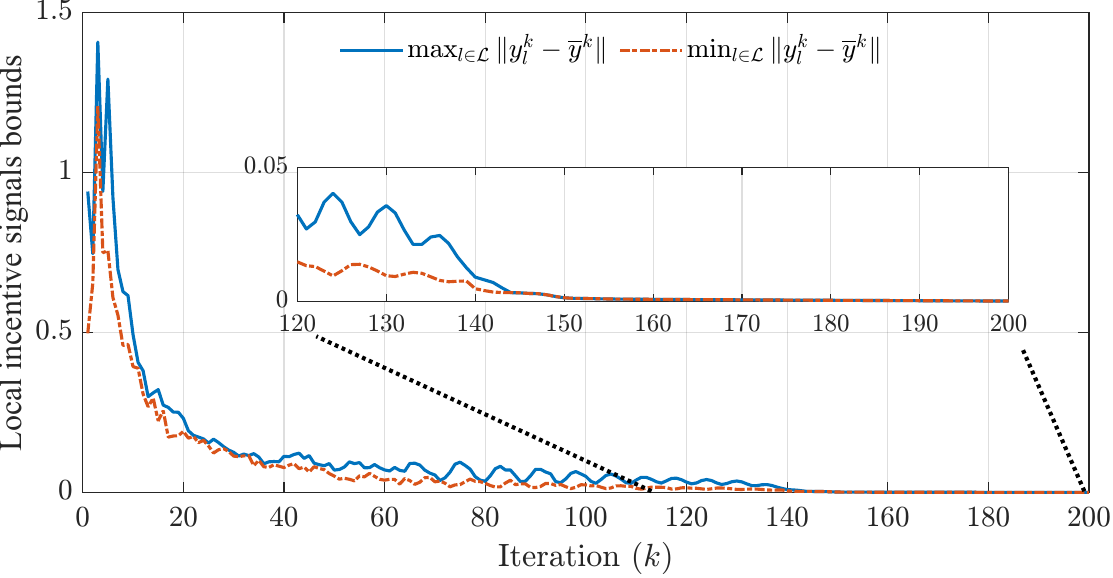}
          \vspace{-.4 cm}
      \caption{Consensus of the local incentive signals.}
      \label{figurelabel2}
      \vspace{-.7 cm}
   \end{figure}

\begin{figure}[thpb]
      \centering
          \includegraphics[scale=0.42]{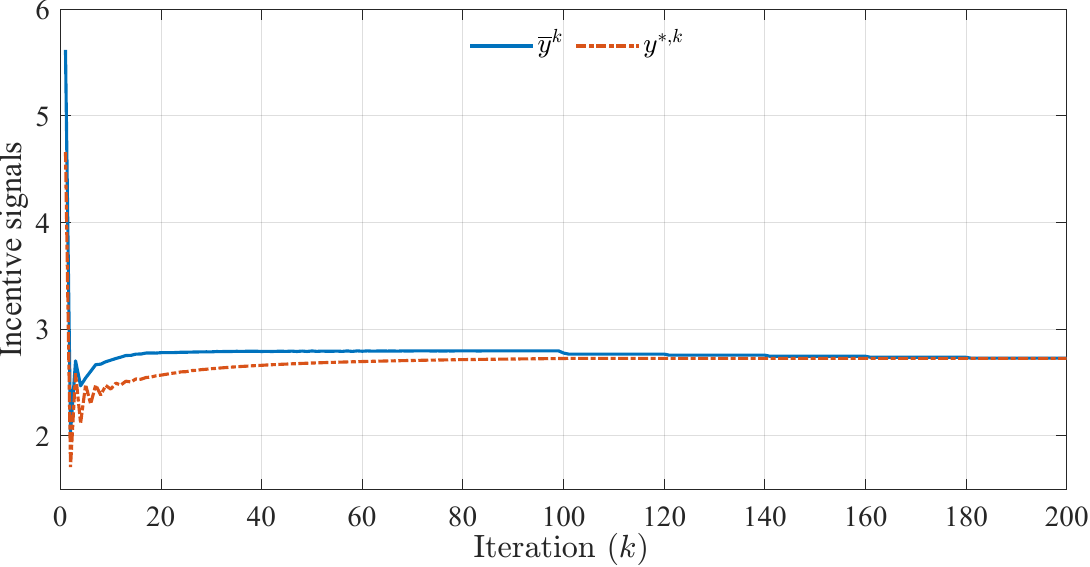}
          \vspace{-.3 cm}
      \caption{Convergence of $\overline{y}^k$ to the global incentive signal $y^{*,k}$ %from Algorithm 1 in \cite{grammatico2017dynamic}
      .  }
      \label{figurelabel3}
      \vspace{-.7 cm}
   \end{figure}

\begin{figure}[thpb]
      \centering \hspace{-0.6 cm}
          \includegraphics[scale=0.37]{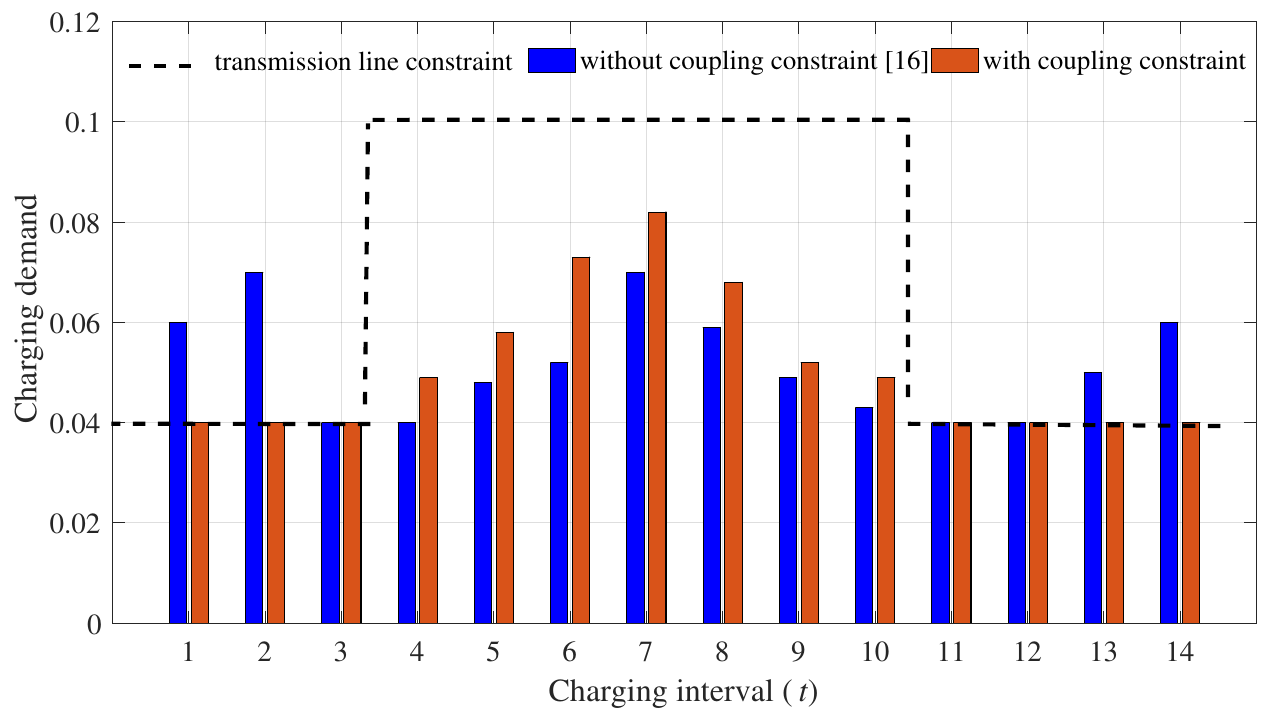}
          \vspace{-.3 cm}
      \caption{Charging demand with and without coupling constraints.}
      \label{figurelabel4}
      \vspace{-.5 cm}
   \end{figure}

We simulate the problem of controlling the charging schedule of EVs with $n=14$ and $L=10$ populations of EVs. We assume the number of EVs in each population equals $10^3$ ($N_l=10^3, l\in\mathcal{L}$). The parameters of (\ref{ex24}-\ref{ex26}) are borrowed from \cite{Ma2015A}. As for battery degradation cost, we set $q_{l,i}=0.004$ and $p_{l,i}=0.075\mathbf{1_{14}}$. Regarding the unit charging price, we set $a=0.038$, $b=0.06$, and $d$ is assumed as in \cite{Ma2015A}. For the individual constraint set \eqref{ex25}, we consider $\underline{x}_{l,i}=0$ and $\overline{x}_{l,i}=0.25$ and $\beta_{l,i}$ is randomly selected from the range $[0.6,1]$. For the coupling constraints, we consider $\overline{s}^t=0.04$ if $t=\{1,2,3,11,12,13,14\}$ and $0.1$, otherwise. By this choice of $\overline{s}^t$, we restrict the total amount of charging demand when the non-EV demand is high. We tune the gains in \eqref{y_lk+1} as $\alpha_k=\frac{1}{k+1}$, design parameter $K$ as $K=0.08I$, and choose $\eta$ based on design choice 3 in \cite{grammatico2017dynamic}. 
For the aforementioned charging schedule problem, Fig. \ref{figurelabel2} shows the minimum and maximum distance between the local incentive signals $y_l^k$ and the average quantity $\overline{y}^k=\frac{1}{L}\sum_{l\in\mathcal{L}}y_l^k$. As shown, the signals $y_l^k$ reach consensus on $\overline{y}^k$.
To show that the sequence $\overline{y}^k$ in \eqref{yb20} is robust in the sense that it converges to the fixed point of mapping $\mathcal{T}$ in \eqref{T}, we compare our algorithm with single population algorithm in \cite{grammatico2017dynamic}.  To have a reasonable comparison, we set parameters of EVs in Algorithm 1 of \cite{grammatico2017dynamic} similar to our case. Figure \ref{figurelabel3} illustrates that average of local aggregates $\overline{y}^k$ in our framework converges to the global fixed point $y^{*}:=\mathrm{col}(\sigma^*,\lambda^*)$ obtained from Algorithm 1 in \cite{grammatico2017dynamic}. From Fig. \ref{figurelabel2} and Fig. \ref{figurelabel3}, we can conclude that all the local incentive signals converge to a global fixed point of the game's mapping. Figure \ref{figurelabel4} compares charging demand in our method with \cite{kebriaei2021multipopulation} which neglects the coupling constraints in the decision-making procedure. As seen in Fig. \ref{figurelabel4}, the charging demand in \cite{kebriaei2021multipopulation} at $t=1,2,13,14$ violates transmission line constraint \eqref{ex26}, which shows inefficiencies of neglecting the coupling constraints.
\vspace{-.3 cm}

\appendices
\section{Design method to determine matrix $K$, mappings $x^{**}(.)$ and $\mathcal{T}(.)$}
Here we are going to discuss how to determine the design matrix $K$, the mapping $x^{**}(.)$ in (6), and the mapping $\mathcal{T}(.)$ in (7). Our main goal in this paper is to seek the multi-population Nash equilibrium for the case where the agents do not have access to signals $\sigma,\lambda$ and the signals are controlled by multi-coordinators. However in this part, we are going to explain how to determine the Nash equilibrium with the cost function (2) and the coupling constraints (1) for the case where the agents have access to the signal $\sigma,\lambda$ and these signals are controlled by a single coordinator, through which we aim to describe the methods to determine design matrix $K$, the mappings $x^{**}(.)$ and $\mathcal{T}(.)$. The procedure to determine matrix $K$, mappings $x^{**}(.)$ and $\mathcal{T}(.)$ has been borrowed from \cite{grammatico2017dynamic}.

By the assumption that $\hat{x}_l=\mathrm{col}(\hat{x}_{l,1},...,\hat{x}_{l,N_l})$ and $\hat{x}=\mathrm{col}(\hat{x}_1,...,\hat{x}_L)$, we define equilibrium for agents with cost function (2) and coupling constraint (1) as $(\hat{x},\hat{\lambda})$. The equilibrium point $(\hat{x},\hat{\lambda})$ is a set of strategies and control vector such that
\begin{enumerate}
    \item The strategy of each agent is optimal assuming the optimality of the strategy of all agents and the control vector, that is
    \begin{equation}
        \hat{x}_{l,i}\in\arg\min_{x_{l,i}\in\mathcal{X}_{l,i}}J_{l,i}(x_{l,i},\frac{1}{L}\sum_{l\in\mathcal{L}}\sum_{i\in\mathcal{N}_l}\delta_{l,i}\hat{x}_{l,i},\hat{\lambda}).
    \end{equation}
    \item For the control vector $\hat{\lambda}$, the coupling constraints are satisfied, that is
    \begin{equation}
        \frac{1}{L}\sum_{l\in\mathcal{L}}\sum_{i\in\mathcal{N}_l}\delta_{l,i}\hat{x}_{l,i}\in\mathcal{C}.
    \end{equation}
\end{enumerate}
To find the equilibrium, We define the agent's best response mapping $x_{l,i}^*:\mathbb{R}^{n}\to\mathcal{X}_{l,i}$ to the incentive signal $v=C\sigma+K\lambda$ as 
\begin{equation}\label{best_response}
    x_{l,i}^*(v):=\arg\min_{u\in\mathcal{X}_{l,i}}f_{l,i}(u)+v^Tu.
\end{equation}
We also define the aggregation mapping $\mathcal{A}:\mathbb{R}^n \times\mathbb{R}^n\rightarrow\frac{1}{L}\sum_{l\in\mathcal{L}}\sum_{i\in\mathcal{N}_l}\delta_{l,i}\mathcal{X}_{l,i}$ as
\begin{equation}
    \mathcal{A}(\sigma,\lambda)=\frac{1}{L}\sum_{l\in\mathcal{L}}\sum_{i\in\mathcal{N}_l}\delta_{l,i}x_{l,i}^{*}(v).
\end{equation}
If $(\hat{\sigma},\hat{\lambda})$ is found such that $\mathcal{A}(\hat{\sigma},\hat{\lambda})=\hat{\sigma}$, for $\hat{\lambda}\in\mathbb{R}^{n}$, then $\hat{x}_{l,i}=x_{l,i}^{*}(C\hat{\sigma}+K\hat{\lambda})$ and $\hat{\lambda}$ is the equilibrium for agents with cost function (2) and the coupling constraint (1) and the above-mentioned condition (1) is satisfied. However, from $\mathcal{A}(\hat{\sigma},\hat{\lambda})=\hat{\sigma}$, we cannot conclude that the coupling constraint (1) is satisfied. Therefore, if there is mapping $x^{**}:\mathbb{R}^{n}\times\mathbb{R}^{n}\rightarrow\mathcal{C}$ such that 
\begin{equation}\label{zz}
    \hat{\sigma}=\mathcal{A}(\hat{\sigma},\hat{\lambda})=x^{**}(\hat{\sigma},\hat{\lambda}),
\end{equation}
then $(\hat{\sigma},\hat{\lambda})$ satisfies both the above-mentioned conditions for the generalized aggregating game. 

Now, we aim to transfer the problem of determining the equilibrium to the problem of finding a zero of an appropriate mapping. $(\hat{\sigma},\hat{\lambda})$ which satisfies \eqref{zz} is equivalent to the zero of mapping $\theta$ which is defined as follows.
\begin{equation}\label{teta}
    \begin{split}
        \Phi(\begin{bmatrix}\sigma\\\lambda
\end{bmatrix})&=\begin{bmatrix}\sigma-\mathcal{A}(\sigma,\lambda)\\\sigma-2\mathcal{A}(\sigma,\lambda)+x^{**}(\sigma,\lambda)
\end{bmatrix}\\&=\begin{bmatrix}
    I_n&0\\I_n&0
\end{bmatrix}\begin{bmatrix}\sigma\\\lambda
\end{bmatrix}-\begin{bmatrix}\mathcal{A}(\sigma,\lambda)\\2\mathcal{A}(\sigma,\lambda)-x^{**}(\sigma,\lambda)
\end{bmatrix}\\&:=(M+\Gamma)(\begin{bmatrix}\sigma\\\lambda
\end{bmatrix})
    \end{split}
\end{equation}
According to [\citenum{Bauschke2011Convex}, Theorem 25.8], an algorithm could be developed to determine the zero of mapping $\Phi$ in \eqref{teta}. To apply Theorem 25.8 in \cite{Bauschke2011Convex}, the linear mapping $M$ should be maximally monotone and $\Gamma$ be $\beta-cocoercive$ in some Hilbert space. Authors in \cite{grammatico2017dynamic} have proved that if matrix $K$ is chosen such that $K\succ0$ and $C+K\succ0$, then the linear mapping $M$ is maximally monotone in $\mathcal{H}_P$, where
\begin{equation}
   P:= \begin{bmatrix}C+2K&-K\\-K&K
\end{bmatrix}.
\end{equation}
Additionally, authors in \cite{grammatico2017dynamic} have proved that for $K\succ0$ and $C+K\succ0$ and the mapping $x^{**}(.)$ defined as follows, the mapping $\Gamma$ is $\beta$-cocoercive in $\mathcal{H}_P$. 
\begin{equation}\label{x0*}
    x^{**}(y):=\arg\min_{z\in\mathcal{C}}\frac{1}{2}z^Tz+(K(\sigma-\lambda))^Tz
\end{equation}

With these design choices, according to Theorem 25.8 in \cite{Bauschke2011Convex}, for any initial condition $(\sigma^0,\lambda^0)$, the following algorithm is converged to the unique zero of mapping $\Phi(.)$.
\begin{equation}\label{upd}
   \begin{bmatrix}\sigma^{k+1}\\\lambda^{k+1}
\end{bmatrix}=(1-\alpha_k)\begin{bmatrix}\sigma^{k}\\\lambda^{k}
\end{bmatrix}+\alpha_k(I+\eta M)^{-1}(\begin{bmatrix}\sigma^{k}\\\lambda^{k}
\end{bmatrix}-\eta\Gamma\left(\begin{bmatrix}\sigma^{k}\\\lambda^{k}
\end{bmatrix}\right)).
\end{equation}
By the assumption that 
\begin{equation}
    \mathcal{T}(\begin{bmatrix}\sigma\\\lambda
\end{bmatrix})=(I+\eta M)^{-1}(\begin{bmatrix}\sigma\\\lambda
\end{bmatrix}-\eta\Gamma\left(\begin{bmatrix}\sigma\\\lambda
\end{bmatrix}\right)),
\end{equation}
we can write \eqref{upd} as follows.
\begin{equation}
\begin{bmatrix}\sigma^{k+1}\\\lambda^{k+1}
\end{bmatrix}=(1-\alpha_k)\begin{bmatrix}\sigma^{k}\\\lambda^{k}
\end{bmatrix}+\alpha_k\mathcal{T}(\begin{bmatrix}\sigma^{k}\\\lambda^{k}
\end{bmatrix})
\end{equation}
Therefore the problem of finding the zero of mapping $\Phi$ is equivalent to that of finding the fixed point of mapping $\mathcal{T}$ (Authors in \cite{grammatico2017dynamic} have proved that for $K\succ0$ and $C+K\succ0$ and the mapping $x^{**}(.)$ in \eqref{x0*}, $\mathcal{T}$ is non-expansive in $\mathcal{H}_P$).
\begin{theorem1}
For $K\succ0$, $C+K\succ0$, and the mapping $x^{**}(.)$ in \eqref{x0*}, $M$ is maximally monotone and $\Gamma$ is $\beta$-cocoercive in Hilbert space $P$, hence, according to [\citenum{Rockafellar1998Variational}, exercise 12.4], $\theta(.)$ is strictly monotone. Therefore, based on [\citenum{Bauschke2011Convex}, Proposition 23.35], the mapping $\theta(.)$ has at most one zero and the equilibrium $(\hat{\sigma},\hat{\lambda})$ is unique.
\end{theorem1}
   
%%%%%%%%%%%%%%%%%%%%%%%%%%%%%%%%%%%%%%%%%%%%
%%%%%%%%%%%%%%%%%%%%%%%%%%%%%%%%%%%%%%%%%%%%%%%%%%%%%%%%%%%%%%%%%%%%%%%%%%%%%%%%
%\pagebrea

\section{Multi-population $\epsilon$-Nash equilibrium}
\newtheorem{theorem6}{Proposition}
\begin{theorem6}
A set of best response strategies to a fixed point of mapping $\mathcal{T}$ in (7) is an $\epsilon$-Nash equilibrium of the game defined by (1) and (2), where $\epsilon$ is uniformly decreasing with the total number of agents. 
\end{theorem6}
\begin{proof}
We assume that $\hat{y}:=\mathrm{col}(\hat{\sigma},\hat{\lambda})$ is a fixed point of mapping $\mathcal{T}$ in (7). Also, we assume that a set of strategies $\hat{x}:=\mathrm{col}(\hat{x}_1,...,\hat{x}_L)$
with $\hat{x}_l=\mathrm{col}(\hat{x}_{l,1},...,\hat{x}_{l,N_l}), l\in\mathcal{L}$ are best responses to the fixed point of the mapping $\mathcal{T}$:
\begin{equation}\label{xhat1}
    \hat{x}_{l,i}=x_{l,i}^{*}(\hat{v}),
\end{equation}
where $x_{l,i}^{*}(.)$ is defined in (6) and 
\begin{equation}\label{vhat}
    \hat{v}=C\hat{\sigma}+K\hat{\lambda},\quad \hat{\sigma}=\frac{1}{L}\sum_{l\in\mathcal{L}}\sum_{i\in\mathcal{N}_l}\delta_{l,i}\hat{x}_{l,i}.
\end{equation}
We also define 
\begin{equation}\label{xti}
    \Tilde{x}_{l,i}=x_{l,i}^{*}(\Tilde{v}),\quad \overbigdot{x}_{l,i}=x_{l,i}^{*}(\overbigdot{v}),
\end{equation}
where 
\begin{equation}\label{vti}
    \begin{split}    &\Tilde{v}=C\Tilde{\sigma}+K\hat{\lambda},\quad 
        \Tilde{\sigma}=\frac{1}{L}(\delta_{l,i}x_{l,i}+\sum_{l\in\mathcal{L}}\sum_{j\neq i}\delta_{l,j}\hat{x}_{l,j})\\
        &\overbigdot{v}=C\overbigdot{\sigma}+K\hat{\lambda},\quad\overbigdot{\sigma}=\frac{1}{L}(\delta_{l,i}\Tilde{x}_{l,i}+\sum_{l\in\mathcal{L}}\sum_{j\neq i}\delta_{l,j}\hat{x}_{l,j}).
    \end{split}
\end{equation}
Then, we can conclude that 
\begin{equation}\label{JJ}
    J_{l,i}(\overbigdot{x}_{l,i},\overbigdot{\sigma},\hat{\lambda})\leq J_{l,i}(\Tilde{x}_{l,i},\overbigdot{\sigma},\hat{\lambda})\leq J_{l,i}(\hat{x}_{l,i},\hat{\sigma},\hat{\lambda}).
\end{equation}
Based on $\eqref{xti}$, the first inequality holds since $\overbigdot{x}_{l,i}$ is the best response to $\overbigdot{v}$, which depends on $\overbigdot{\sigma}$ and $\hat{\lambda}$. The second inequality holds as the best response to $\hat{v}$ in \eqref{vhat} leads to a higher cost than the case each agent considers its strategy effect on the aggregate term, thus
\begin{equation}\label{JJ2}
 J_{l,i}(\hat{x}_{l,i},\hat{\sigma},\hat{\lambda})-J_{l,i}(\Tilde{x}_{l,i},\overbigdot{\sigma},\hat{\lambda})\leq J_{l,i}(\hat{x}_{l,i},\hat{\sigma},\hat{\lambda})-J_{l,i}(\overbigdot{x}_{l,i},\overbigdot{\sigma},\hat{\lambda}).  
\end{equation}
Based on \eqref{JJ2} and the assumption that $J_{l,i},l\in\mathcal{L}, i\in\mathcal{N}_l$ in (2) are Lipschitz continuous, we have
\begin{equation}\label{JJJ}
   J_{l,i}(\hat{x}_{l,i},\hat{\sigma},\hat{\lambda})-J_{l,i}(\overbigdot{x}_{l,i},\overbigdot{\sigma},\hat{\lambda})\leq \alpha(\|\overbigdot{\sigma}-\hat{\sigma}\|+\|\hat{x}_{l,i}-\overbigdot{x}_{l,i}\|).
\end{equation}
We also assume that the best response mapping $x_{l,i}^{*}(.)$ in $\mathrm{(4)}$ is non-expansive, so according to \eqref{xhat1} and \eqref{xti}
\begin{equation}\label{leqq}
    \|\hat{x}_{l,i}-\overbigdot{x}_{l,i}\|\leq\|\overbigdot{v}-\hat{v}\|.
\end{equation}
In view of $\hat{v}$ in \eqref{vhat} and  $\overbigdot{v}$ in \eqref{vti}, $\|\overbigdot{v}-\hat{v}\|\leq\|C\|\|\overbigdot{\sigma}-\hat{\sigma}\|$. Then, by using \eqref{leqq} and \eqref{JJJ}, we have
\begin{equation}\label{ll}
    J_{l,i}(\hat{x}_{l,i},\hat{\sigma},\hat{\lambda})-J_{l,i}(\overbigdot{x}_{l,i},\overbigdot{\sigma},\hat{\lambda})\leq \alpha(\|C\|+1)\|\overbigdot{\sigma}-\hat{\sigma}\|.
\end{equation}
Based on \eqref{vti} and \eqref{vhat}, we have
\begin{equation}\label{***}
    \|\overbigdot{\sigma}-\hat{\sigma}\|=\frac{\delta_{l,i}}{L}\|\Tilde{x}_{l,i}-\hat{x}_{l,i}\|.
\end{equation}
Since $\mathcal{X}_{l,i}$ is compact, 
\begin{equation}\label{****}
    \|\Tilde{x}_{l,i}-\hat{x}_{l,i}\|\leq\|\Tilde{x}_{l,i}\|+\|\hat{x}_{l,i}\|\leq2R
\end{equation}
By using \eqref{JJ2}, (\ref{ll}-\ref{****}) and assuming that $\delta_{l,i}$ is uniformly bounded, i.e. $\delta_{l,i}\leq\overline{\delta}/N_l$, we obtain
\begin{equation}
\begin{split}
        J_{l,i}(\hat{x}_{l,i},\hat{\sigma},\hat{\lambda})-J_{l,i}(\tilde{x}_{l,i},\overbigdot{\sigma},\hat{\lambda})&\leq\frac{2\alpha(\|C\|+1)R\overline{\delta}}{LN_l}\\&\leq\frac{2\alpha(\|C\|+1)R\overline{\delta}}{N}:=\epsilon.
\end{split}
\end{equation}
Thus, Algorithm 1 converges to a set of strategies which are $\epsilon$-Nash, where $\epsilon$ is uniformly decreasing with a total number of agents, $N$.    
\end{proof}
   \begin{figure}[thpb]
      \centering
          \includegraphics[scale=0.4]{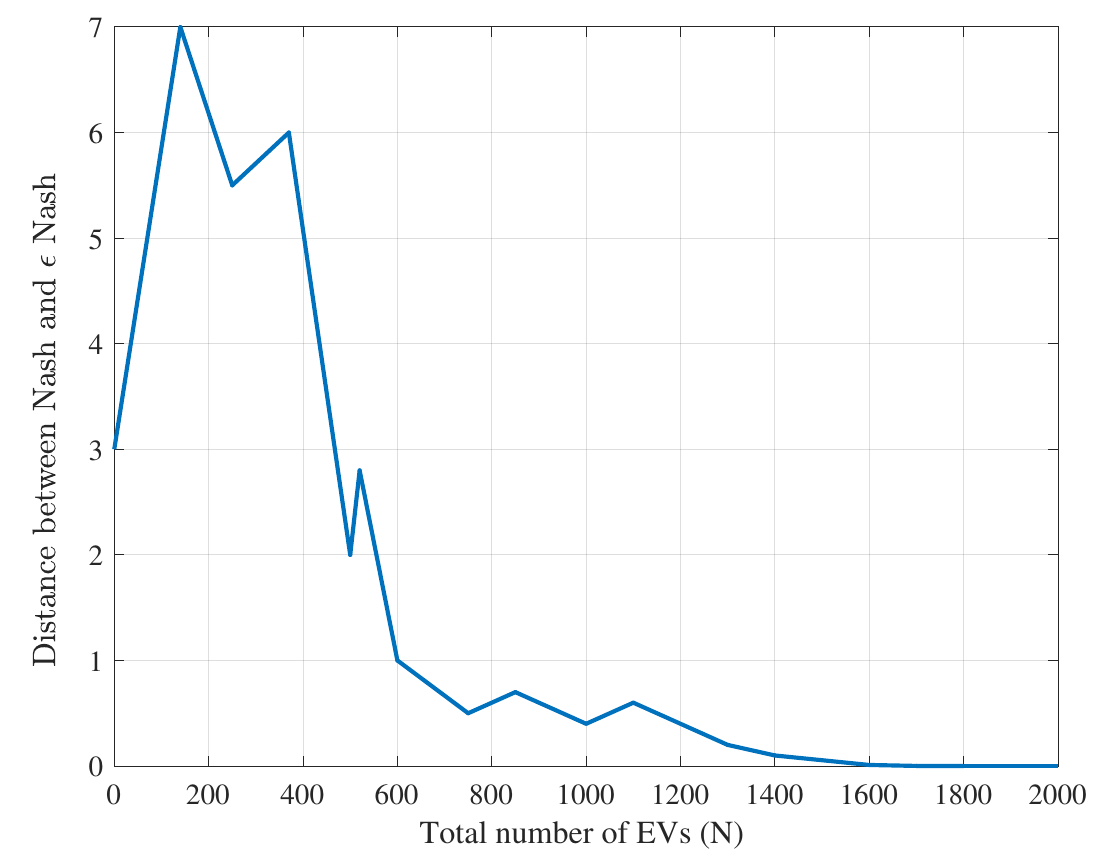}
      \caption{Distance between $\epsilon$-Nash and Nash equilibrium as a function of total numbers of EVs. }
      \label{figurelabel5}
   \end{figure}
Figure \ref{figurelabel5} illustrates the distance between $\epsilon$-Nash and Nash equilibrium parametric on the total number of EVs. As shown, $\epsilon$ converges to zero with increasing total numbers of EVs. 

%%%%%%%%%%%%%%%%%%%%%%%%%%%%%%%%%%%%%%%%%%%%%%%%%%%%%%%%%%%%%%%%%%%%%%%%%%%%%%%%%%%%%%%%%%%%%%%%%%%%%%%%%%%%%%%%%%%%%%%%%%%%%%%%%%%

\bibliographystyle{IEEEtran}
\bibliography{Ref}
\end{document}